\documentclass[fleqn,11pt]{wlscirep}
\usepackage{caption}
\usepackage{subcaption}
\usepackage{stackengine}
\usepackage{graphicx}
\usepackage{float}
\usepackage{xr}
\title{Propulsion and controlled steering of magnetic nanohelices}
\usepackage{setspace}
\author[1,*]{Maria Michiko Alcanzare}
\author[2,3]{Mikko Karttunen}
\author[1,4]{Tapio Ala-Nissila}
\affil[1]{COMP CoE at the Department of Applied Physics, Aalto University School of Science, P.O. Box 11000, 
FIN-00076 Aalto, Espoo, Finland}
\affil[2]{Department of Mathematics and Computer Science \& Institute for Complex Molecular Systems, Eindhoven University of Technology, P.O.Box 513, MetaForum 5600 MB Netherlands}
\affil[3]{Department of Chemistry \& Applied Mathematics, Western University, 1151 Richmond Street, London, Ontario, Canada N6A 5B7}
\affil[4]{Department of Physics, Box 1843, Brown University, Providence, Rhode Island 02912-1843, USA}
\affil[*]{maria.alcanzare@aalto.fi}
\affil[+]{the authors contributed equally to this work}
\newcommand{\vect}[1]{\mathbf{#1}}

\begin{abstract}

Externally controlled motion of micro and nanomotors in a fluid environment constitutes a promising tool in biosensing, targeted delivery and environmental remediation. In particular, recent experiments have demonstrated that fuel-free propulsion can be achieved through the application of external magnetic fields on magnetic helically shaped structures. The magnetic interaction between helices and the rotating field induces a torque that rotates and propels them via the coupled rotational-translational motion. Recent works have shown that there exist certain optimal geometries of helical shapes for propulsion. However, experiments show that controlled motion remains a challenge at the nanoscale due to Brownian motion that interferes with the deterministic motion and makes it difficult to achieve controlled steering. In the present work we employ quantitatively accurate simulation methodology to design a setup for which magnetic nanohelices of 30 nm in radius, with and without cargo, can be accurately propelled and steered in the presence of thermal fluctuations. In particular, we demonstrate fast transport of such nanomotors and devise protocols in manipulating external fields to achieve directionally controlled steering at biologically relevant temperatures.

\end{abstract}
\begin{document}

\flushbottom
\maketitle

\thispagestyle{empty}

\section*{Introduction}

The development of artificial nano and micromotors that can be controlled accurately and precisely in spatial and temporal scales has attracted rapidly increasing interest.
They 
are typically catalytically driven \cite{paxton2004catalytic,paxton2005motility,wang2006bipolar,Laocharoensuk2008,ke2010motion,mirkovic2010nanolocomotion,solovev2012,ma2015catalytic,safdar2015,kline2005catalytic,Wang2008,Kagan2010,wang2010motion} or magnetically propelled  \cite{ghosh2009controlled,tottori2012magnetic,Nelson2014,Gao2012,schamel2014nanopropellers,Walker,sanchez2011,medina2015cellular,magdanz2016dynamic,bechinger2016active,dreyfus2005microscopic}. 
One of the key issues for such nanomotors 
to be useful for practical
purposes is remote motion control and maneuverability.
This has been demonstrated where thermal fluctuations are not important such as in the transport, guidance, and release of cells
\cite{sanchez2011,medina2015cellular,magdanz2016dynamic}, single-cell targeted delivery \cite{Nelson2014} and oil droplet capture \cite{zhao2011external,Guix2012}. 
Catalytically driven micromotors rely on chemical gradients for self-propulsion, and
motion control is made possible through externally manipulated magnetic fields \cite{kline2005catalytic,Wang2008,Kagan2010,wang2010motion}.  Catalysis-propelled particles require fuel such as hydrogen peroxide, acidic or alkaline solutions which poses a challenge in their biocompatibility for \textit{in vivo} applications \cite{Wang2014}. This challenge can be alleviated by creating catalytically driven micromotors that require low concentrations of fuel without compromising their propulsion speeds \cite{Gao11} or using micromotors that propel in biocompatible solutions \cite{Gaowaterdriven2012,mou2013self}. Magnetic propellers, on the other hand, are suitable candidates for \textit{in vivo} applications since they do not require fuel and most soft biological tissues have very low magnetic susceptibilities \cite{Schenck}. Such propellers have been shown to be driven and elaborately maneuvered, and can perform complex tasks such as cargo fetching and delivery
\cite{tottori2012magnetic,Gao2012,schamel2014nanopropellers,medina2015cellular}. Their propulsion speeds can be remotely controlled by tuning the field frequencies. 

Scaling magnetic propellers down to the nanosize presents a major challenge for controlled motion and maneuverability because of thermal fluctuations\cite{mirkovic2010nanolocomotion,wang2010motion,schamel2014nanopropellers};  thermal Brownian motion can severely alter the direction of motion and interfere with the propulsion. Schamel {\it et al.} \cite{schamel2014nanopropellers} circumvented the challenge in nanoscale steering by driving the nanopropellers at low frequencies in gels and highly viscous media to damp out thermal fluctuations. In water where biology occurs, however, no propulsion was observed as Brownian effects dominated the propulsion \cite{schamel2014nanopropellers}.

\begin{figure}[h]
\centering
\includegraphics[width=0.55\textwidth]{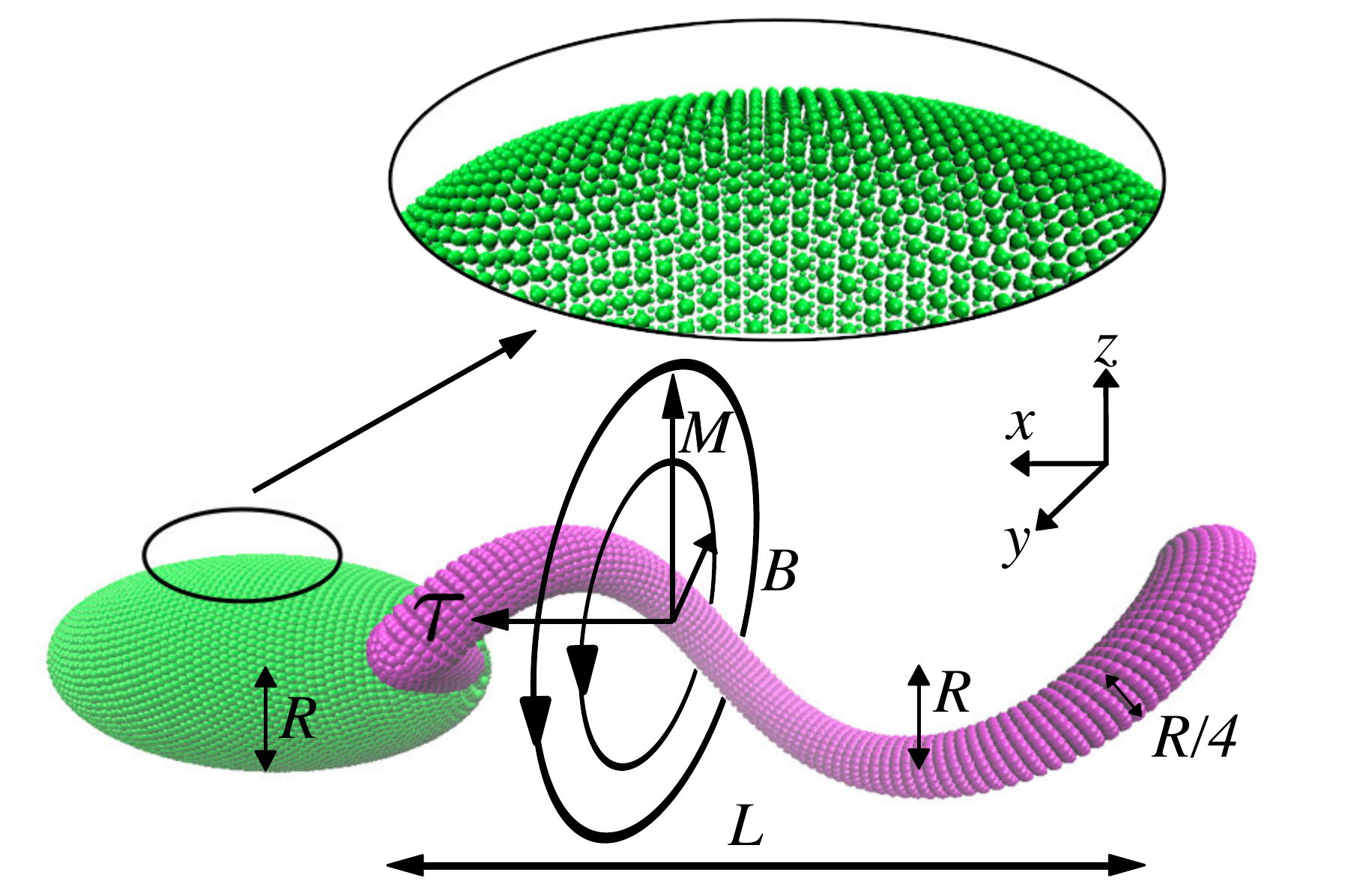}
\caption{Node representation of the extended bodies in the fluctuating lattice-Boltzmann - Molecular dynamics method \cite{Ollila2011}. The optimal helix has $N=1.25$ turns and a turn length $L$
to radius $R$ ratio
of $L/R \approx 6$, with $R=30$ nm for nanoscale helices. The optional cargo is a prolate spheroid, which minimizes hydrodynamic drag \cite{zabarankin2010three}
and whose minor axis equals $R$ 
(See Supporting Information for more details).
The magnetic moment of the helix is fixed perpendicular to the long axis such that torque is along the long axis of propulsion, as shown schematically in the figure.}
\label{figure1}
\end{figure}

In our previous work \cite{Alcanzare2017}, we studied the shape dependence of rotation and propulsion of helically shaped magnetic nanoparticles at constant torque. 
We found that the maximum propulsion is achieved by balancing competing requirements for rotational stability and minimizing viscous drag. Of various helical shapes, we 
found that the helix with a circular cross-section and number of turns $N=1.25$ and length to radius ratio $L/R\approx6$ exhibits maximal propulsion velocity (cf. Fig. \ref{figure1} and Fig. \ref{SI-Fig1}).
Furthermore, for well-defined propulsion and chiral separation
in the presence of Brownian motion, the P\'eclet number must be larger than about $50$.
In the present work, we extend our previous work \cite{Alcanzare2017} to the experimentally relevant case where the helix has a fixed magnetic moment and 
it is driven by an external magnetic field with varying frequency $f$. In particular, 
we show how steered propulsion can be optimized
as a function of the
magnetic interaction and $f$. We also consider the case where the helix functions as a cargo-carrying nanomotor, and present a protocol for changing the field for
controlled directional steering
of the nanopropellers. 
We use the recently developed fluctuating lattice-Boltzmann method
\cite{Ollila2011,Ollilasiam,mackay2013coupling,mackay2013hydrodynamic,citelammps} to quantitatively model the fluid environment and the nanopropellers.
Details about the method are provided in the Supporting Information (SI). 

\subsection*{Coupled translational and rotational motion by magnetic field}

At the scales of micron and less, micro-organisms navigate in water based fluids at very low Reynolds numbers ${\rm Re} \ll 1$. 
They have developed, through billions of years of evolution, 
helically shaped appendages or flagella with locomotion as the primary function \cite{Jarrell2008}. Artifical micropropellers have a similar shape as the flagella
such that as the propeller rotates in the fluid, it pushes the fluid along the direction of its easy axis of propulsion. At low Re,  
this motion is described by the generalized resistance tensor:
\begin{equation}
\label{eq1}
\left(\begin{matrix}
\vect{F} \\ \vect{\tau}
\end{matrix}\right)
=
\left( \begin{matrix}  \vect\xi^\text{TT} & \vect\xi^\text{TR} \\ \vect\xi^\text{RT} & \vect\xi^\text{RR} \\
\end{matrix} \right)
\left(\begin{matrix}
\vect v \\ \vect\omega
\end{matrix}\right),
\end{equation}
where $\vect F, \vect \tau, \vect v, \vect \omega$ and $\vect\xi^{\alpha\beta}$ are the applied force, applied torque, translational velocity, angular velocity, 
and the friction matrix of the particle, respectively \cite{purcell1977life,Lauga-Powers}. Bodies with spherical symmetry have zero non-diagonal elements in their 
resistance tensor. Screw-shaped bodies have helicoidal symmetry and 
non-zero off-diagonal terms in Eq. \ref{eq1}.
This means that the translational and the rotational motion are coupled \textit{i.e.} rotational motion will result in 
translational motion and vice versa. Rotation of a magnetic helix 
can be induced
by a torque that is generated by the interaction of its magnetic moment $\vect{M}$ with an external magnetic field $\vect{B}$. As indicated in
Fig. \ref{figure1}, here we set $\vect{M}$ to be uniform and perpendicular to the helix's easy axis of rotation. An external field  is applied and the torque on the helix
is given by $\vect{\tau}=\vect{M}\times\vect{B}$ (see Fig. \ref{figure1}). The field frequencies used in the simulations are $f = 10-10000$ kHz. 
The state-of-the-art nanopropeller's magnetic moment, which were driven in highly viscous media, is estimated to be $2\times10^{-14}$ emu \cite{schamel2014nanopropellers}. 
For $MB=(1.5-2.5)\times10^{-18}$ Nm in the simulations and using the experimental value of the magnetic moment,
the magnetic field strengths required would be $75-125$ mT. The fastest micropropellers (of 100-150 nm radius) had a propulsion speed of $\sim40$ $\mu$m/s in water at 150 Hz and 5 mT \cite{ghosh2009controlled}.  

\section*{Results}

\subsection*{Dependence of Propulsion on Frequency}

\begin{figure}[h]
\centering
\begin{subfigure}{0.40\textwidth}
  \centering
  \includegraphics[width=0.85\textwidth]{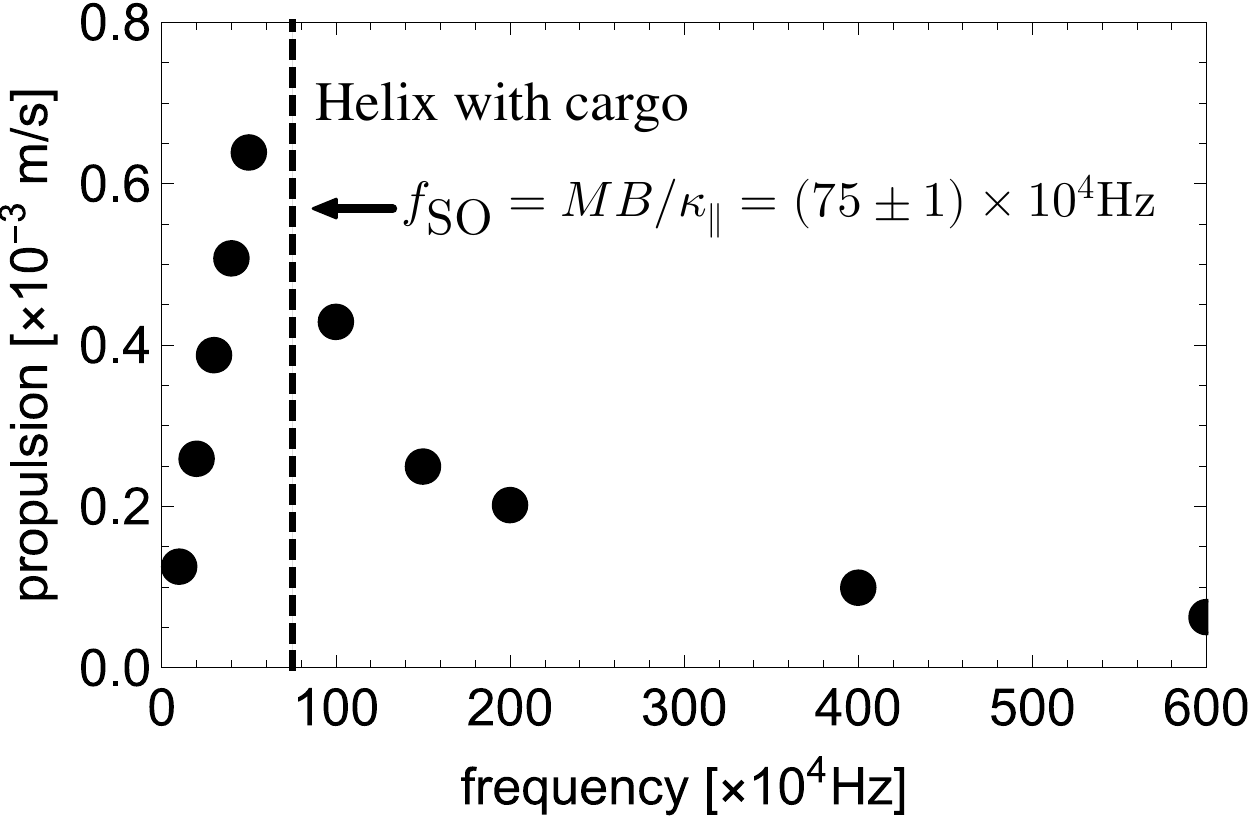}
  \caption{}
  \label{fig2a}
\end{subfigure}
\begin{subfigure}{0.40\textwidth}
  \centering
  \includegraphics[width=0.85\textwidth]{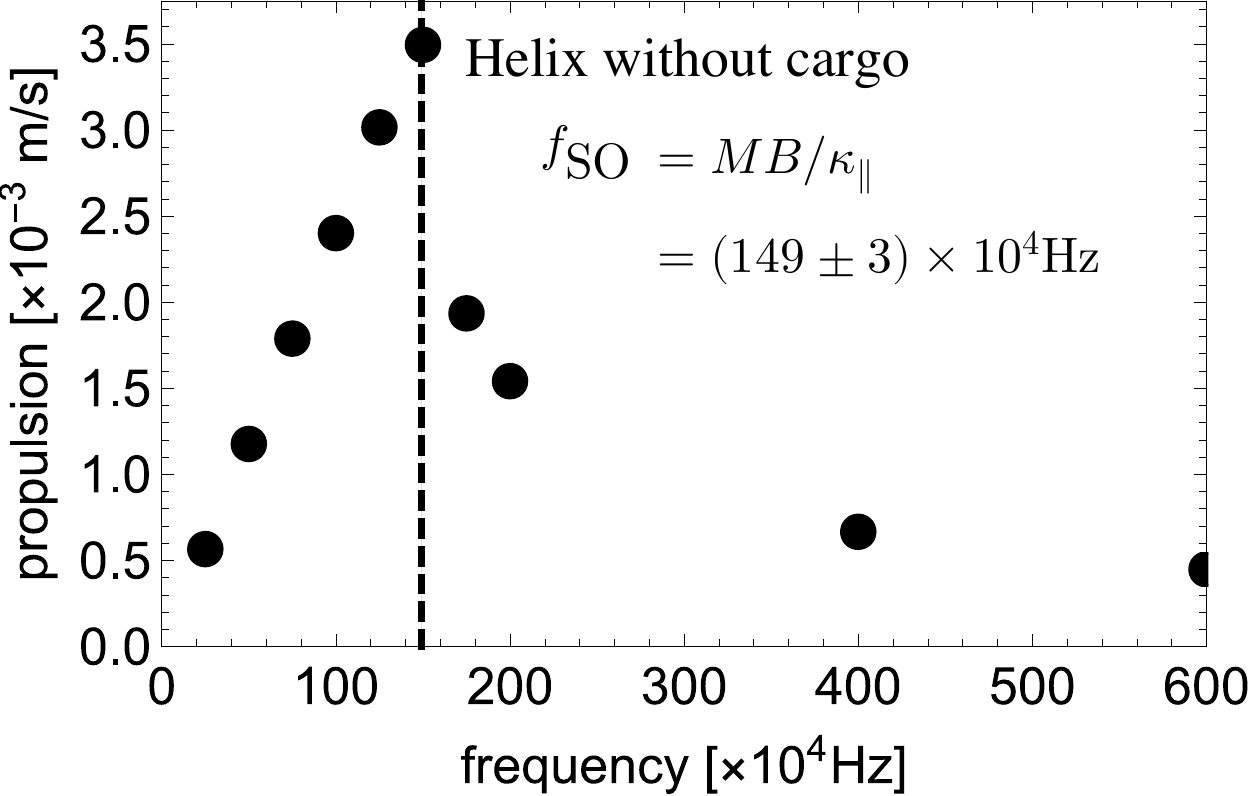}
  \caption{}
  \label{fig2b}
\end{subfigure}%
\caption{Propulsion velocities of a helix with a helical diameter of 30 nm driven at $MB=2.0\times10^{-18}$ Nm. The step-out frequencies are indicated by the blue dashed lines. 
Thermal fluctuations are not included here. 
}
\label{fig2}
\end{figure}

In the limit where Brownian motion can be neglected (${\rm Pe} \gg 1$),
the total net forces and torques on the helix are due to its magnetic interaction with the external field and the hydrodynamic drag. 
For propulsion, an external magnetic field of constant amplitude is rotated along a plane with frequency $f$. The direction of propulsion is perpendicular to the plane of the rotating field (cf. Fig. \ref{figure1}).
Both the magnitude of the applied torque and the hydrodynamic drag increase from zero up to $\kappa_\parallel f$ for frequencies that are less than $MB/\kappa_\parallel$, 
where $\kappa_\parallel$ is the viscous rotational drag coefficient; Fig. \ref{SI-Fig3a} shows the response of the hydrodynamic drag on the rotation of the helix. 
Once the helix has achieved an angular frequency of rotation that is equal to that of the magnetic field, the difference between the applied torque and the 
magnitude of the hydrodynamic drag goes to zero, and the phase difference between the magnetic moment and the magnetic field becomes constant (Fig. \ref{SI-Fig3b}).

Increasing the frequency of the external field enhances propulsion due to the translational-rotational coupling (Eq. \ref{eq1}), as shown in Fig. \ref{fig2}.  
The propulsion is linear with the frequency since the torque attains a constant value at steady state. However,
when the frequency is greater than a threshold step-out value $f_{\rm SO}=MB/\kappa_\parallel$, the helix eventually lags the magnetic field ((Fig. \ref{SI-Fig4})). 
For frequencies less than the step-out frequency, the phase difference reaches a constant value below $\pi/2$ and the resulting external torque also attains a 
constant value (Fig. \ref{SI-Fig5}). For field frequencies greater than the step-out frequency, the phase difference ``steps-out" of $\pi/2$ 
and consequently varies periodically as the process of the magnetic moment lagging the magnetic field and the magnetic field catching up to the magnetic moment repeats itself. 
Accordingly, the external torque varies periodically 
for $f>f_{\rm SO}$ as shown in Fig. \ref{SI-Fig5}. This is reflected in the strong reduction in propulsion
beyond $f_{\rm SO}$ (Fig. \ref{fig2}). We find that the step-out frequency of the helix with an attached cargo (Fig. \ref{SI-Fig5b}) (a prolate spheroid which
minimizes the Stokes drag coefficient) is 
smaller by about a factor of two than the step-out frequency of the helix without the cargo (Fig. \ref{SI-Fig5a}). This is consistent with the fact that 
the helix with an attached cargo has a greater viscous drag coefficient due to the additional surface area of the cargo than a helix without an attached cargo
(the drag coefficients are given in Table \ref{SI-Table2}).


\subsection*{P\'eclet numbers}

\begin{figure}[h!]
\centering
\includegraphics[width=0.45\textwidth]{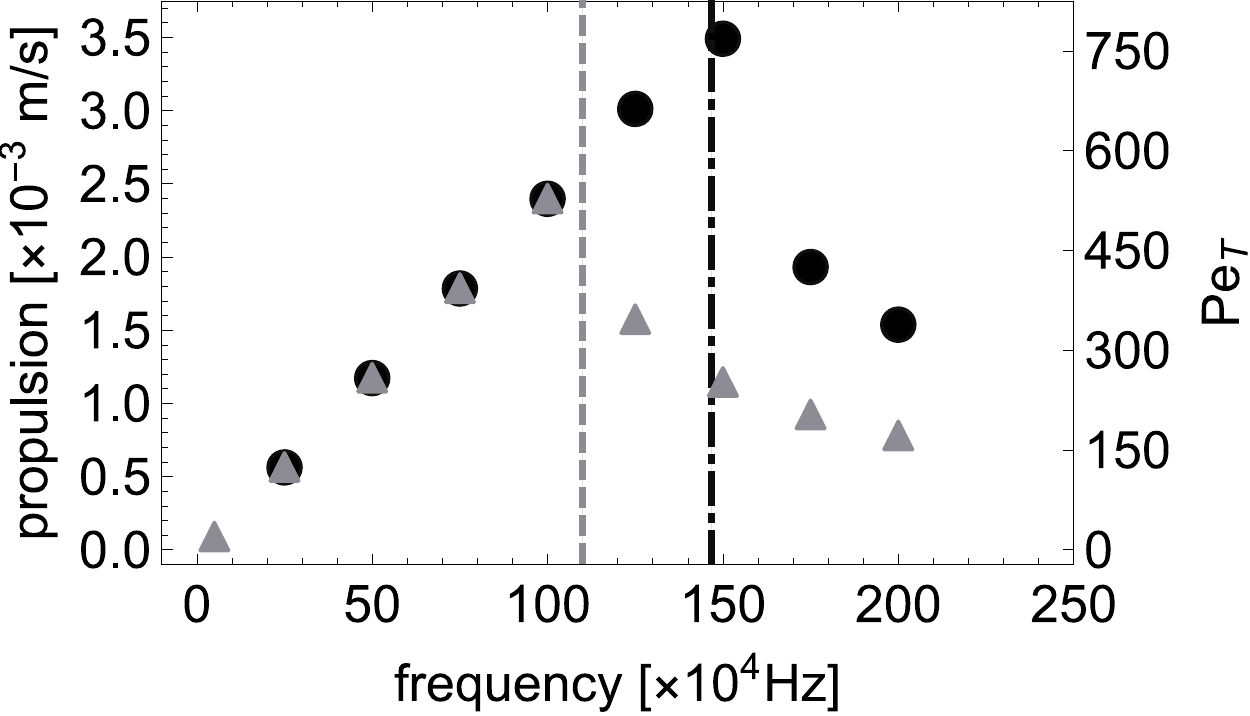}
\caption{Propulsion velocity of a helix without an attached cargo with thermal fluctuations at $T=300$ K. Circles and triangles correspond to a maximum torque of 
$1.5\times10^{-18}$Nm and $2.0\times10^{-18}$ Nm, respectively. The dashed lines mark the corresponding step-out frequencies $f_{\rm SO}=(109\pm2)\times10^4$ Hz 
and $f_{\rm SO}=(149\pm3)\times10^4$ Hz, which increase linearly with $MB$. 
The error bars are smaller than the markers (see Figs. \ref{SI-Fig8} and \ref{SI-Fig9}). Representative displacement plots at various field frequencies are in Fig. \ref{SI-Fig7}.}
\label{fig3}
\end{figure}

The importance of Brownian motion to propulsion can be quantified by considering the P\'eclet number, which is the ratio between the 
diffusive and the advective time scales. A P\'eclet number greater than one means that the advective motion dominates over thermal effects. 
For translational motion, the P\'eclet number is given by $\text{Pe}_{\rm T}=vL/D_{\rm T}$ and for rotational motion $\text{Pe}_{\rm R}=\Omega/D_{\rm R}$, 
where $D$, $v$ and $\Omega$, and $L$ refer to the diffusion coefficient, the propulsion velocity, angular velocity, and length of the helix, 
respectively \cite{schamel2014nanopropellers}. In our previous study we found that nanohelices could be reliably propelled for Pe$_{\rm T}\gtrsim50$ \cite{Alcanzare2017}.
For the $30$ nm helices here, the minimum frequencies for $\text{Pe}_{\rm R}>1$ are at $(1558 \pm 5)$ and $(3095 \pm 60)$ Hz with and
without cargo, respectively (cf. Table \ref{SI-Table1}). The frequencies used in the simulations thus result in P\'eclet numbers 
that are much larger than unity for both with and without cargo as shown in Fig. \ref{fig3}, where we plot the propulsion velocities and the corresponding
translational P\'eclet numbers for two different values of $MB$ with Brownian motion at $T=300$ K. We note that
the highest propulsion speed of the nanohelices that were driven at $MB=2.0\times10^{-18}$ Nm and $f=149\times10^4$ Hz is $\sim3.5\times10^{-3}$ m/s which is $\sim90$ times greater than the propulsion of the fastest experimental micropropellers that were driven at $150$ Hz ($\sim40\times10^{-6}$ m/s) \cite{ghosh2009controlled,pak2011high}. 

\subsection*{Directional steering of nanohelices}

In addition to steering propellers along a straight path, directional maneuverability is a crucial aspect for the application of helices as nanomotors and cargo carriers.
For the magnetic propellers, reversing the direction of motion is a matter of switching the direction of the rotation of the external magnetic field. Alterations in the direction of motion may be done by changing the direction of the torques or changing the perpendicular plane of the rotating field. 
Here we demonstrate steering both with and without thermal fluctuations and at high P\'eclet numbers (${\rm Pe_T}=370$ and 526 for 700 kHz and 1000 kHz, respectively) such that the external 
driving is significantly greater than Brownian motion.

Two types of steering were tested corresponding to high and low values of $MB$. The aim at high $MB$ steering is to avoid the scenario where the magnetic moment 
of the helix steps out of the $\pi/2$ phase with the magnetic field. 
High $MB$ steering subjects the helix to high external torques during the turn. Low $MB$ turning, on the other hand, results
to turns for which the helix steps out of the $\pi/2$ phase for a short time and simultaneously rotates the helix in the opposite direction of the
field rotation until it synchronizes back into the rotation of the magnetic field.

\begin{figure}[H]
\centering
High $MB$ steering \\
\begin{minipage}{0.46\textwidth}
\begin{subfigure}{0.485\textwidth}
  \centering
  \includegraphics[width=1\textwidth]{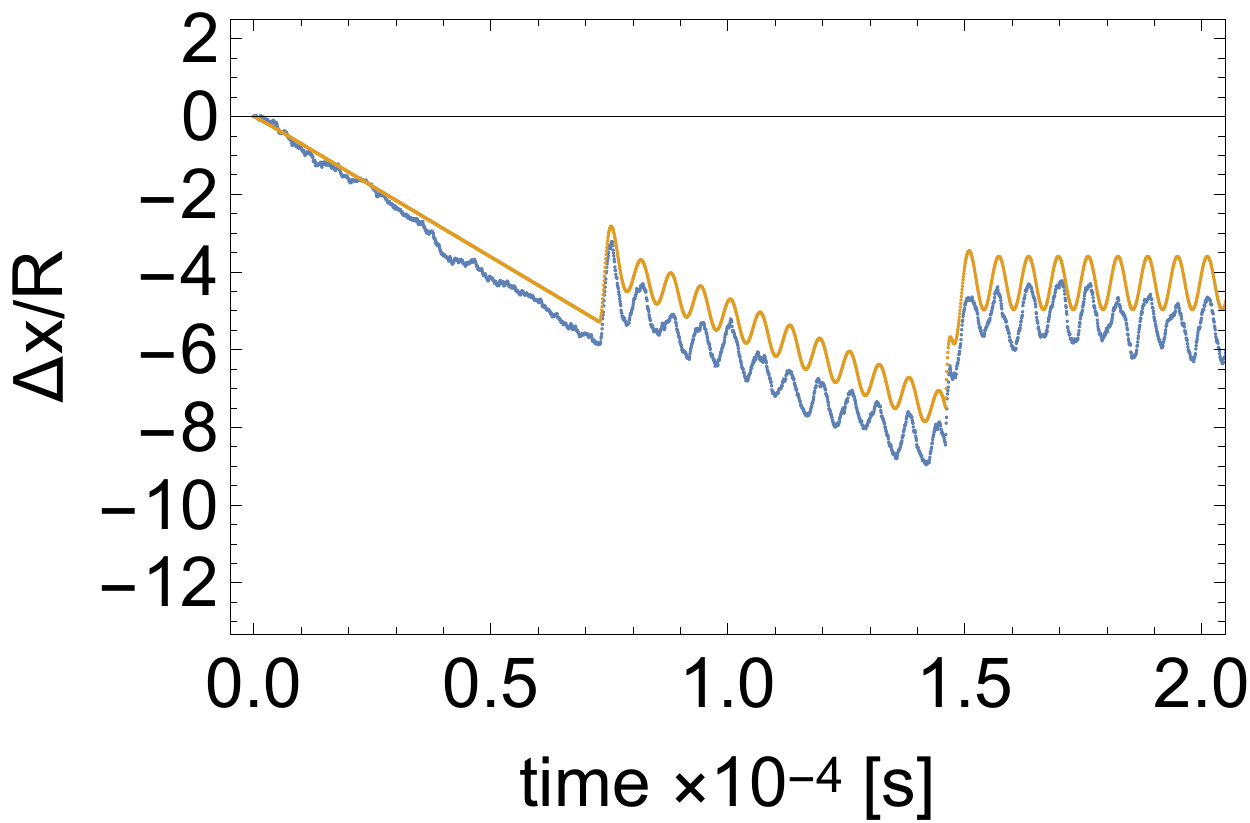}
  \caption{}
  \label{fig4}
\end{subfigure}%
\begin{subfigure}{0.485\textwidth}
  \centering
  \includegraphics[width=1\textwidth]{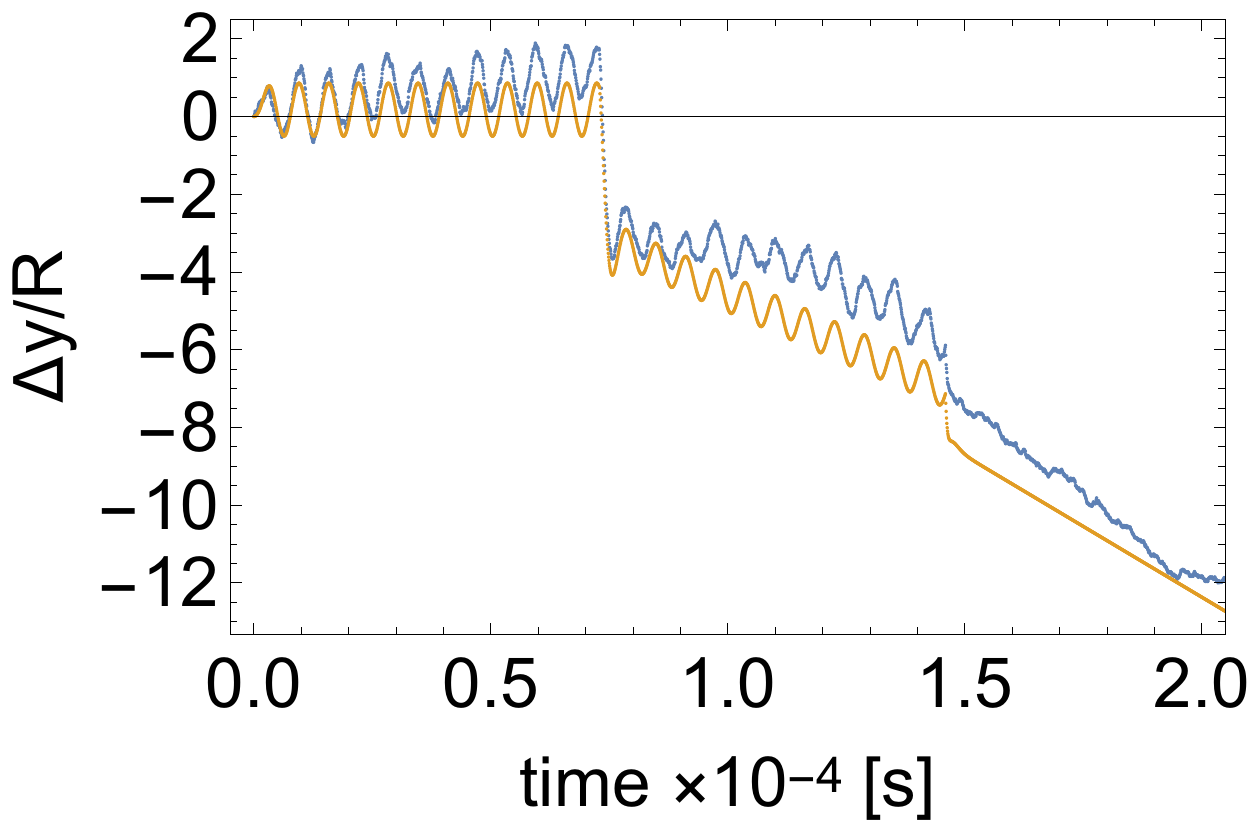}
  \caption{}
  \label{fig:steeringsub2}
\end{subfigure}\\
\begin{subfigure}{0.485\textwidth}
  \centering
  \includegraphics[width=1\textwidth]{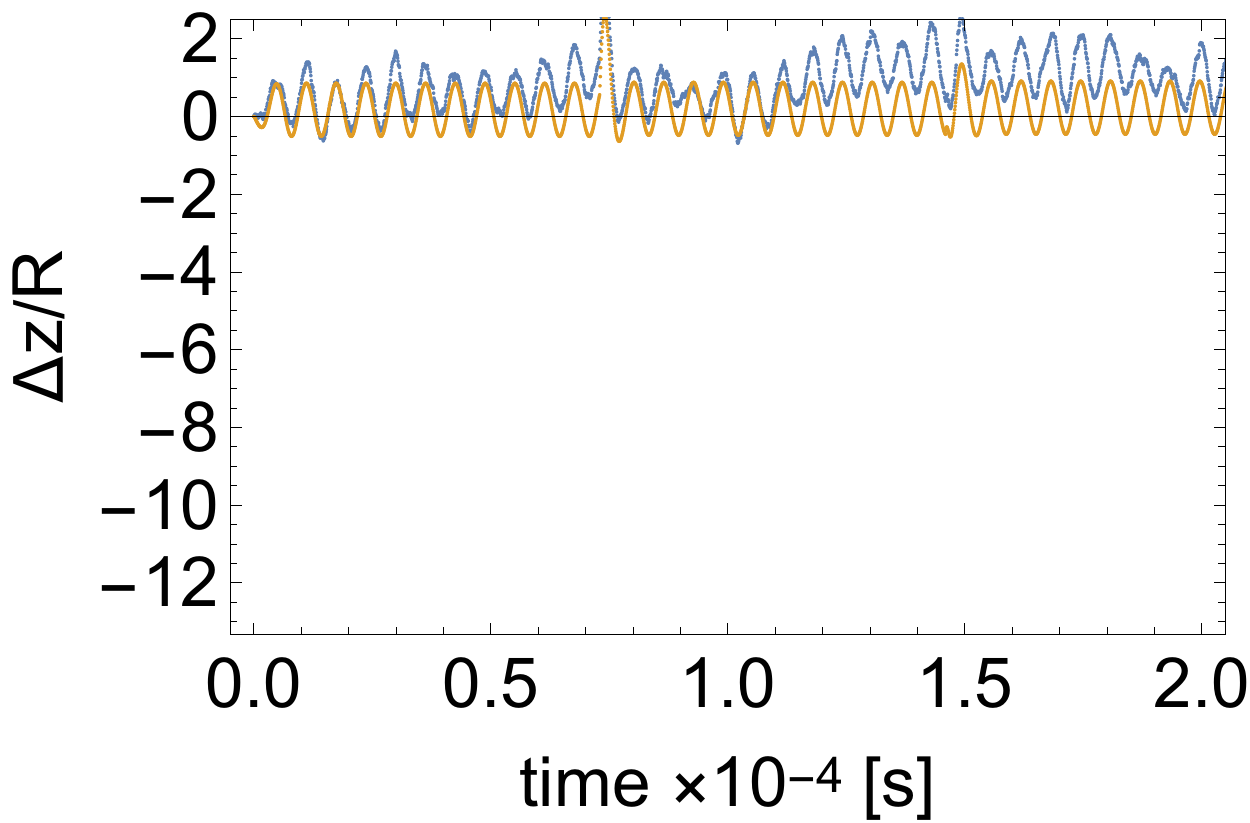}
  \caption{}
  \label{fig:steeringsub3}
\end{subfigure}
\begin{subfigure}{0.485\textwidth}
  \centering
  \includegraphics[width=1\textwidth]{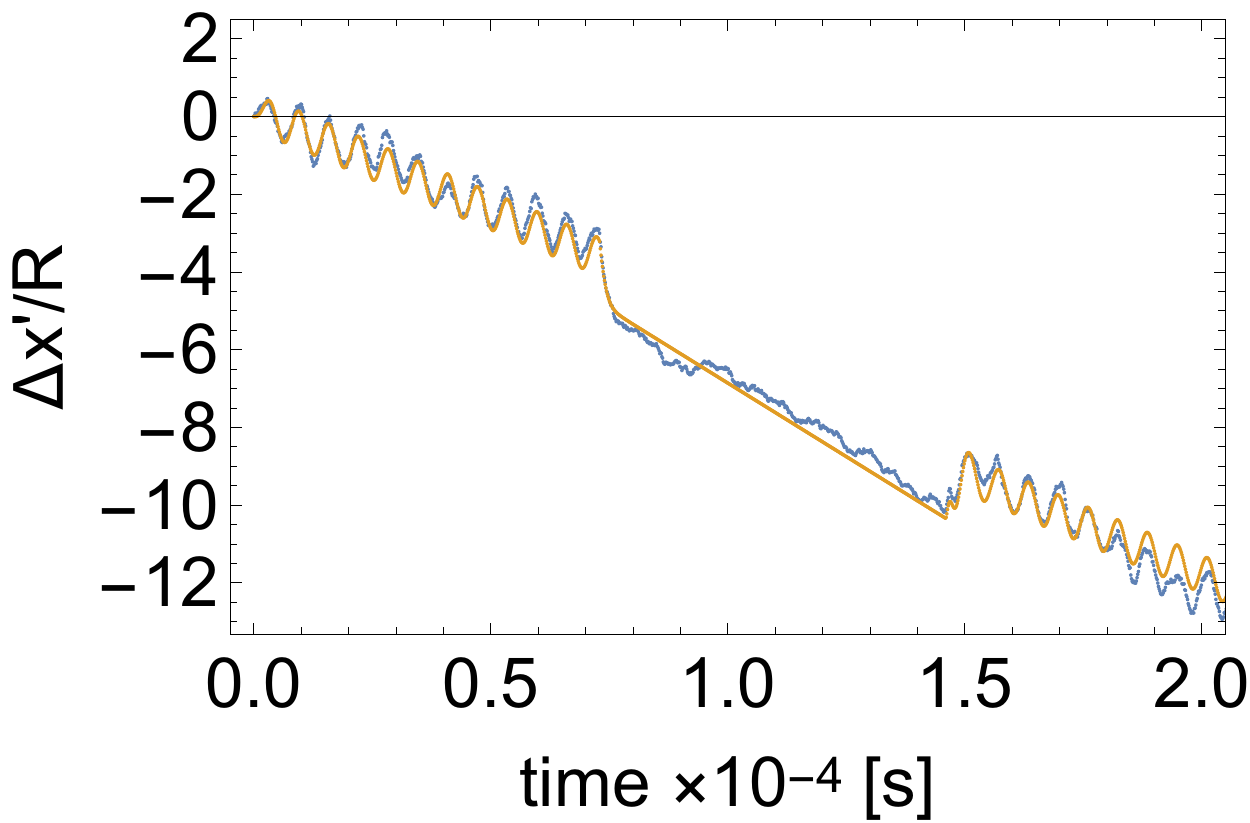}
  \caption{ }
  \label{fig:steeringsub4}
\end{subfigure}
\end{minipage}
\begin{minipage}{0.40\textwidth}
\begin{subfigure}{0.9\textwidth}
  \centering
  \includegraphics[width=0.88\textwidth]{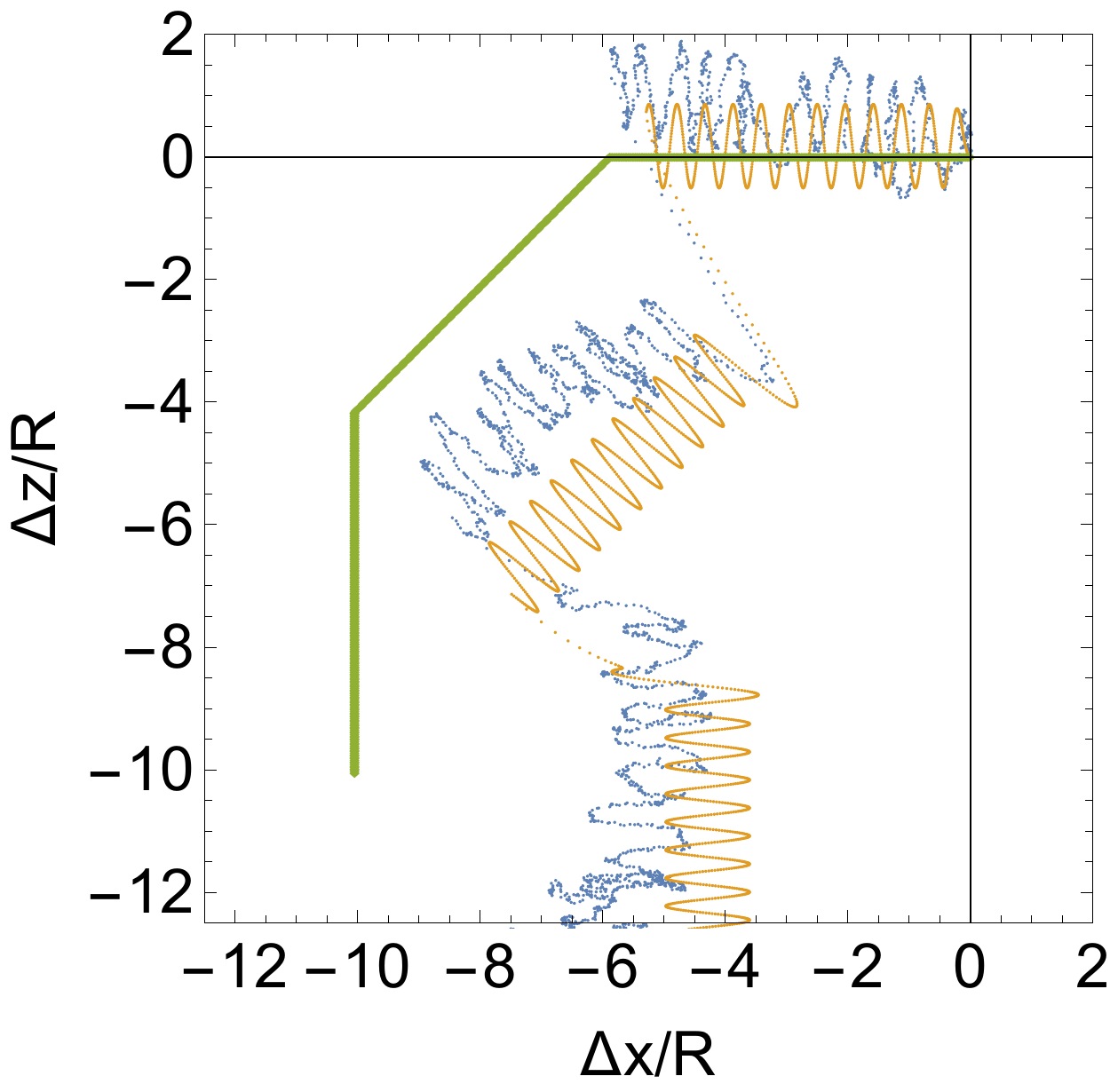}
  \caption{ }
  \label{fig:steeringparapetricfail}
\end{subfigure}
\end{minipage} \vspace{0.2cm}
 \\ 
\hrulefill\\
Low $MB$ steering
\\ \vspace{0.2cm}
  \begin{subfigure}{0.40\textwidth}
  \centering
    \includegraphics[height=0.160\textheight]{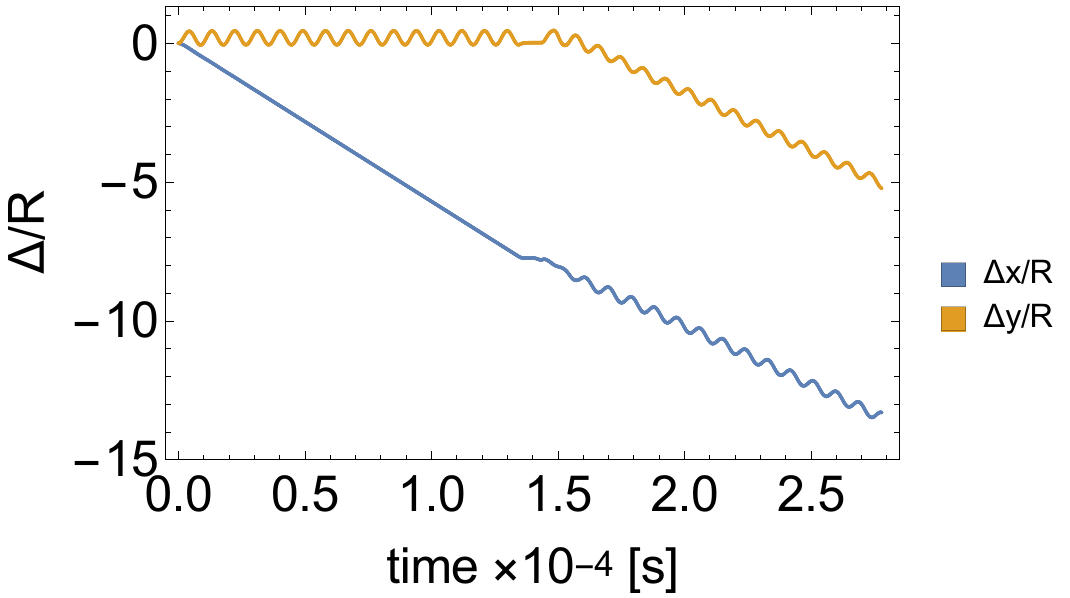}
    \caption{ }
  \label{fig:steeringsub6}
  \end{subfigure}
  \begin{subfigure}{0.40\textwidth}
  \centering
  \includegraphics[height=0.160\textheight]{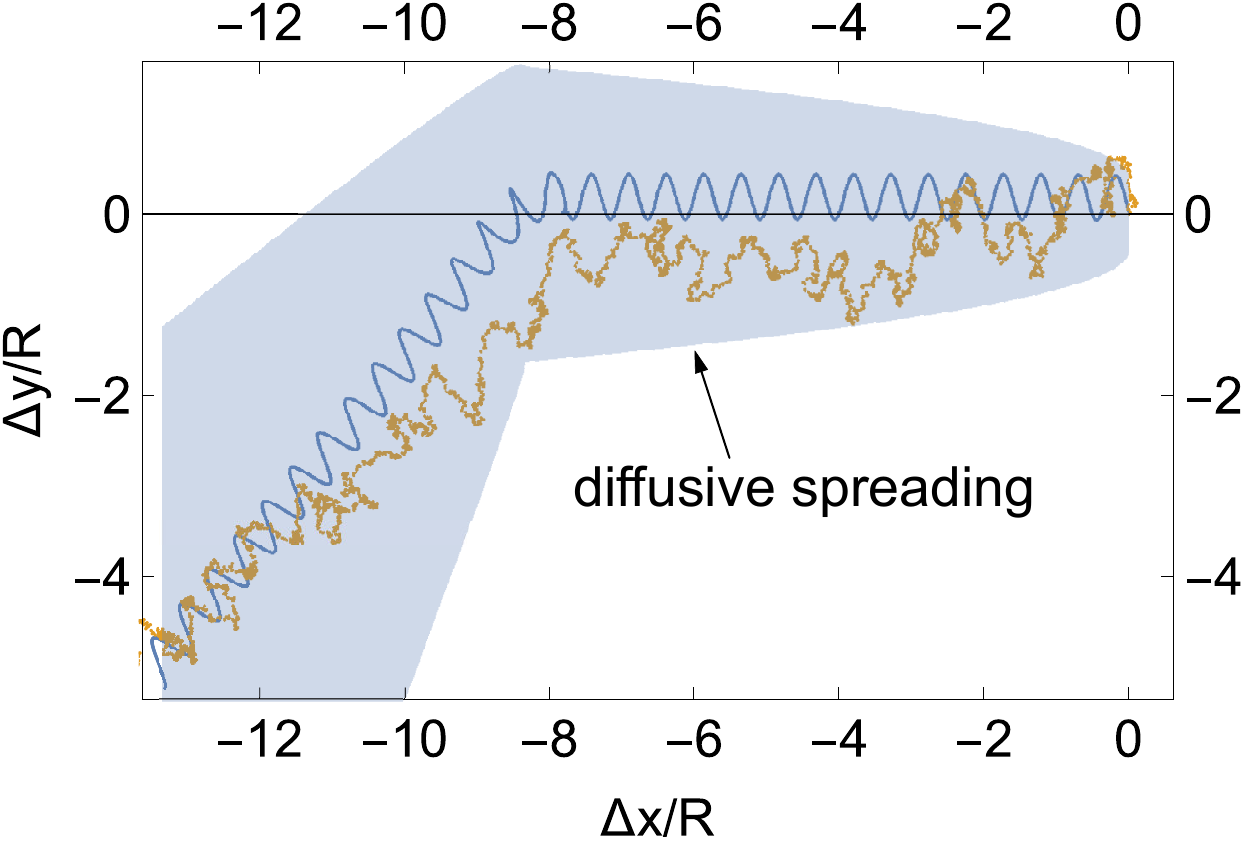}
  \caption{ }
  \label{fig:steeringsub7}
  \end{subfigure}\\
  \begin{subfigure}{0.40\textwidth}
  \centering
    \includegraphics[height=0.160\textheight]{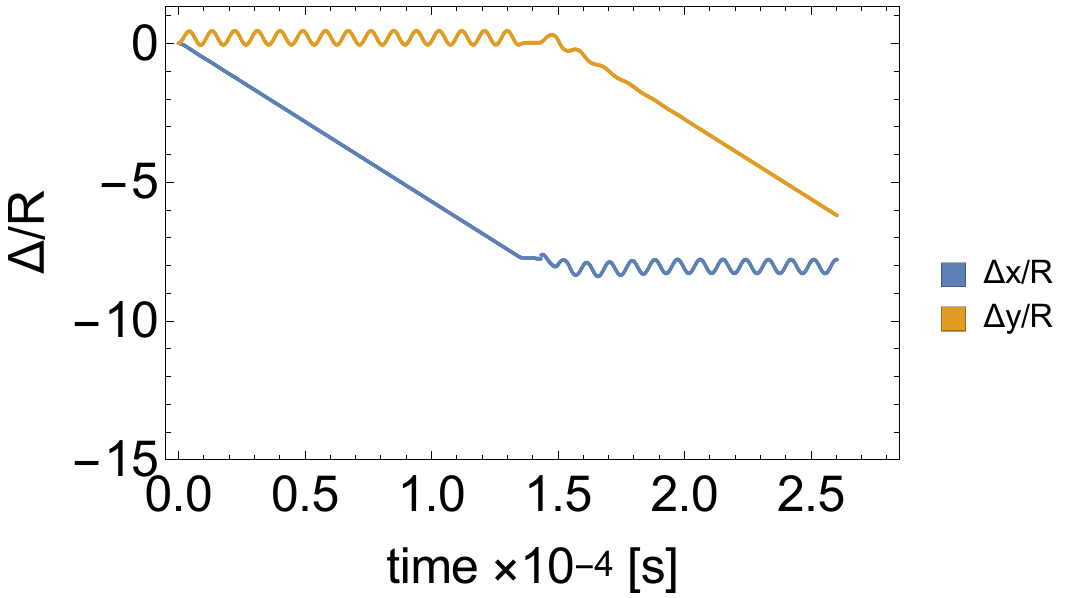}
    \caption{ }
  \label{fig:steeringsub8}
  \end{subfigure}
  \begin{subfigure}{0.40\textwidth}
  \centering
  \includegraphics[height=0.160\textheight]{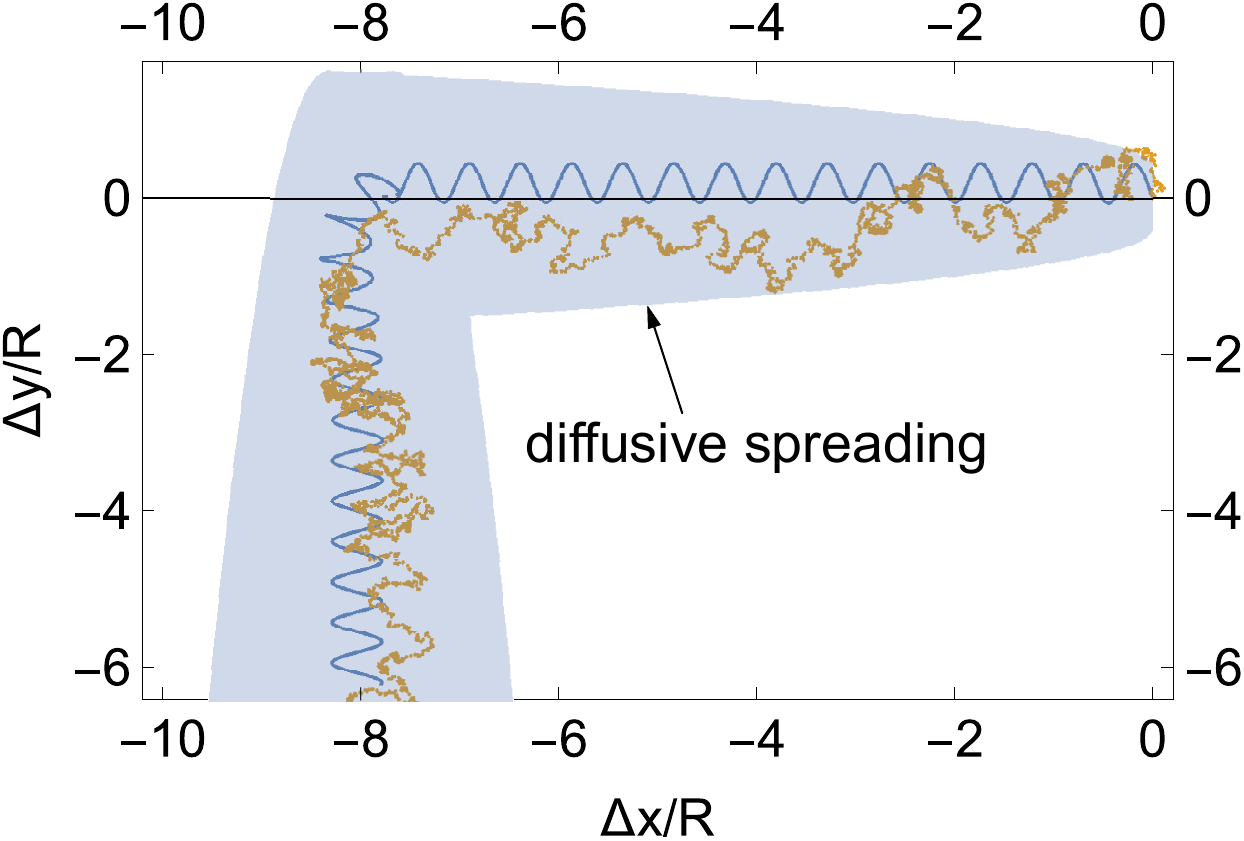}
  \caption{ }
  \label{fig:steeringsub9}
  \end{subfigure}
\caption{Figures \textbf{(a-e)} illustrate directional high $MB$ steering of a nanohelix at $f=1000$ kHz with $MB=10\times10^{-18}$ Nm, where the perpendicular plane of the 
magnetic field is instantaneously changed during the turns. The steering protocol has three stages: propulsion in the $x$-direction and two consecutive 
rotations by $45^\circ$ on the $xy$ plane (see text for details). {\bf(e)} The blue, orange and green trajectories are the paths for steering with thermal fluctuations
corresponding to $T=300$ K, without thermal
fluctuations and the intended path, respectively. 
Panels \textbf{(a-c)} are the normalized displacements $\Delta/R$ along the normal coordinates, panel \textbf{(d)} is the $x'$ displacement in a 
transformed coordinate that is rotated by $45^\circ$ in the $z$-direction, and panel \textbf{(e)} is a parametric plot of the trajectory. Due to the jumps, the intended path is not accurately followed.
Panels \textbf{(f-i)} illustrate low $MB$ steering of a nanohelix at $f=700$ kHz with $MB=1.5\times10^{-18}$ Nm, where the changes in the magnetic 
field during the turns are done gradually to prevent asynchronization of the propeller with the field (see text for details).
Panels {\bf(f)} and {\bf(h)} show the $\Delta x/R$ and the $\Delta y/R$ displacements for $45^\circ$ and $90^\circ$  turns, and  panels {\bf(g)} and {\bf(i)} 
show the corresponding parametric trajectories. The intended path is now faithfully followed. The shaded region indicates diffusive spreading due to Brownian
motion at $T=300$ K.}
\label{fig:steeringtrajectory}
\end{figure}

Figures \ref{fig4}-\ref{fig:steeringsub4} demonstrate the case of high $MB$ steering. 
Two $45^\circ$ turns were performed at $1000$ kHz and $MB=10\times10^{-18}$ Nm. The steering protocol is as follows: first, the nanohelix is 
propelled for $2.5\times10^6$ time steps by applying an external torque in the $-x$-direction. Then the direction of the torque is instantaneously 
changed towards $-(\hat{x}+\hat{y})$ for the same amount of time steps (Fig. \ref{SI-Fig6}). In the final part, the helix is propelled in the $-y$-direction.
Small oscillations in the displacement along the long axis are present as shown in Figs. \ref{fig:steeringsub2} and \ref{fig:steeringsub3} 
for $0 < t <0.7\times10^{-4}$ s when the helix is propelled along the $-\hat{x}$, and for $1.4\times10^{-5} < t < 2.1\times10^{-4}$ s when 
the helix is propelled along $-\hat{y}$. These oscillations 
in the displacement are not observed in the direction of the propulsion as shown in Fig. \ref{fig4} and in the transformed 
coordinates $x'$ rotated by $45^\circ$ from the $x$-axis when the helix is turned by $45^\circ$ at $t=0.7\times10^{-5}$ s (Fig. \ref{fig:steeringsub4}). 
These turns were done at high $MB$ such that the magnetic moment of the helix does not step out of the $\pi/2$ phase with the magnetic field. 
Despite this, the large torque causes the helix to turn with a sudden motion causing a sharp jump in the 
displacement during the turns (at $t=0.7\times10^{-5}$ s and at $t=1.4\times10^{-5}$ s). These jumps are 
more prominent in the parametric plot of the trajectory in Fig. \ref{fig:steeringparapetricfail} where it can also be seen that the intended trajectory is not closely followed by the helix.

In contrast, for low steering at $MB=1.5\times10^{-18}$ Nm with the steering protocol described above,
the magnetic moment of the helix momentarily steps
out of the $\pi/2$ phase difference. During this stage, it rotates in a direction opposite to the magnetic field until its rotational motion is synchronized 
with that of the field. This makes controlled directional steering
difficult to achieve 
as demonstrated in Fig. \ref{SI-Fig10} with abrupt, sharp $90^\circ$ turns.

To overcome this difficulty and to ensure 
that the helix is kept synchronized with the magnetic field, gradual changes in the direction of the field must be made. 
%
%
To this end, we have devised the following steering protocol. At the beginning the helix is first synchronized with the field 
(by propelling it for $15$ rotations of the magnetic field here). The actual turning begins by fixing the magnetic field in 
a given direction for a period of time $\delta t=T_1$ to allow the helix to realign with the fixed field as it prepares to be steered. 
To achieve a turn, the field is then rotated in small increments that add up to $\theta$. After this the field direction is fixed again for $\delta t=T_3$. 
The time interval for which the gradual changes in the field are done must be $f^{-1} < \delta t < \tau_\text{relax}$, 
where $\tau_\text{relax} \approx \delta R^2/D_T$ is the time scale to traverse a distance of $\delta R \le L$ by diffusion. 
In this manner asynchronization is avoided and successful
$\theta=45^\circ$ and $\theta=90^\circ$ turns are demonstrated for $MB=1.5\times10^{-18}$ Nm in 
Figs. \ref{fig:steeringsub6}-\ref{fig:steeringsub9}. Without Brownian motion (blue trajectories in Figs. \ref{fig:steeringsub7} and \ref{fig:steeringsub9}) the intended trajectory
is faithfully followed due to the lack of sudden jumps. In the presence of thermal fluctuations, the trajectory follows the intended path within the region of diffusive spreading which can be mitigated by optimizing the propulsion velocity. 
In the simulations we chose $T_1=4.4\times10^{-6}$ s, $T_2=2.9\times10^{-6}$ s and $T_2=1.5\times10^{-6}$ s.

The shaded regions in the parametric plots (Figs. \ref{fig:steeringsub7} and \ref{fig:steeringsub9}) represent the regions of diffusional spreading caused by Brownian motion at $T=300$ K as 
obtained from the mean square displacement measurements. This inherent spreading,
of width $\delta s$, that is orthogonal to the direction of propulsion is
controlled by the Brownian tracer diffusion coefficient of the helix during the straight segments of the path by $\delta s \approx \sqrt{D_{\rm T} t_s}$,
where $t_s$ is the duration of time from the start of the propulsion. 
Thus, diffusional spreading can be mitigated by increasing the
propulsion velocity (for a given path length), or reducing $D_{\rm T}$
by lowering the temperature, increasing the propeller size
and increasing the viscosity of the fluid \cite{Alcanzare2017}.

\begin{figure}[h!]
\centering
\begin{minipage}{0.75\textwidth}
\begin{subfigure}{0.5\textwidth}
  \centering
  \includegraphics[width=0.95\textwidth]{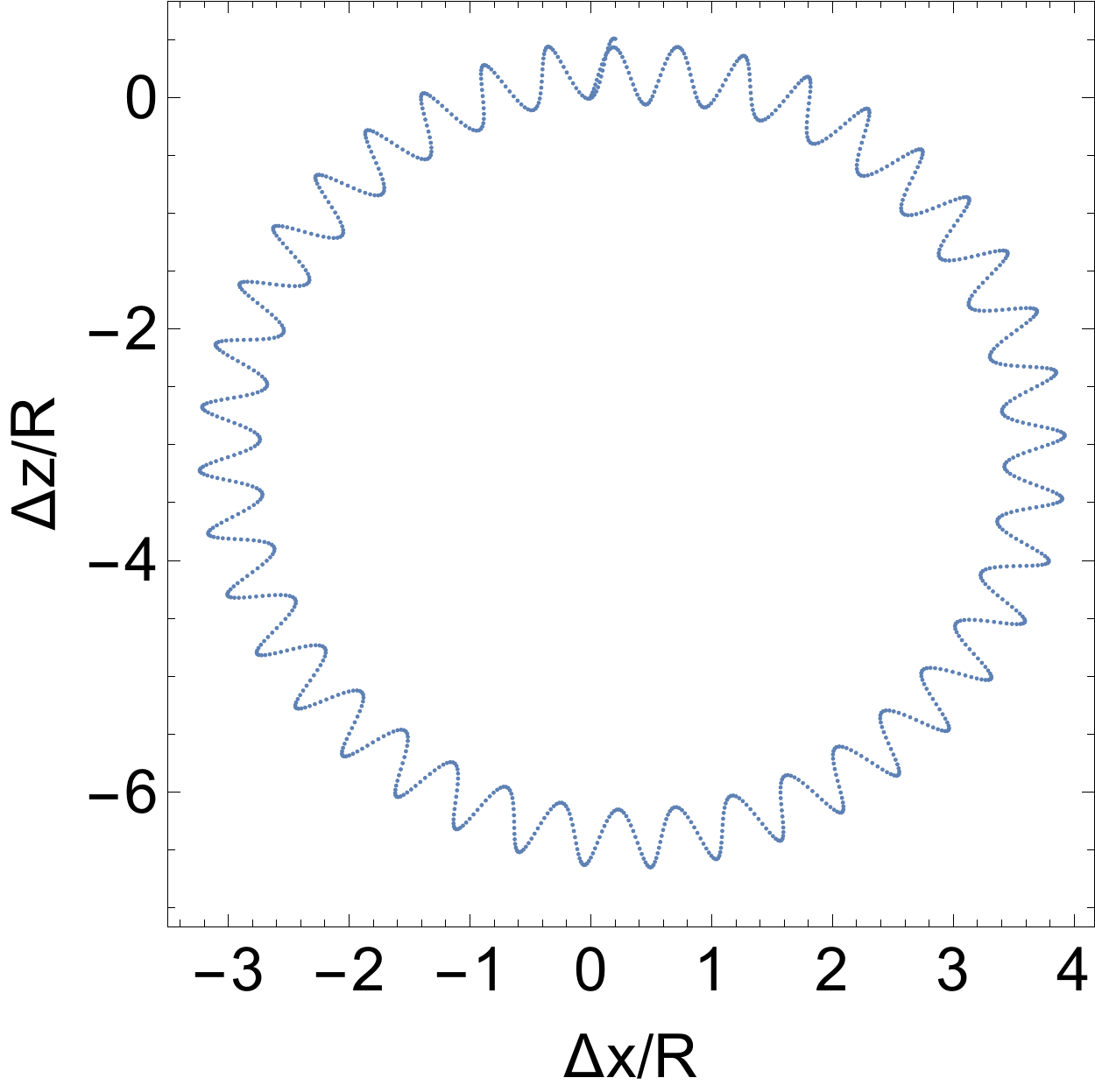}
  \caption{circular turn}
  \label{fig:1sub1}
\end{subfigure}%
\begin{subfigure}{0.5\textwidth}
  \centering
  \includegraphics[width=0.95\textwidth]{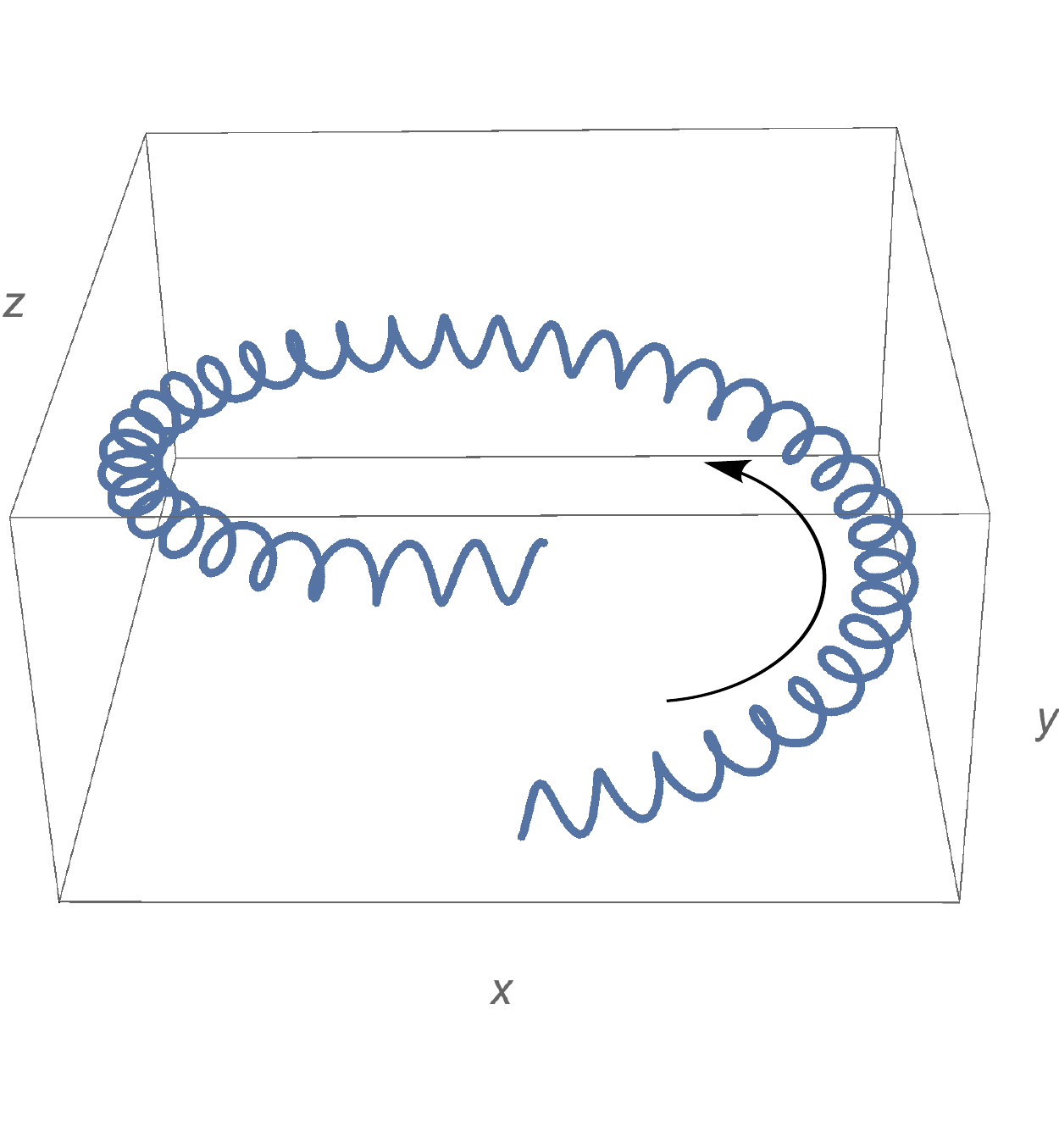}
  \caption{3D plot of the circular turn}
  \label{fig:2sub2}
\end{subfigure}
\end{minipage}
\begin{minipage}{0.24\textwidth}
\begin{subfigure}{1\textwidth}
\includegraphics[width=1\textwidth]{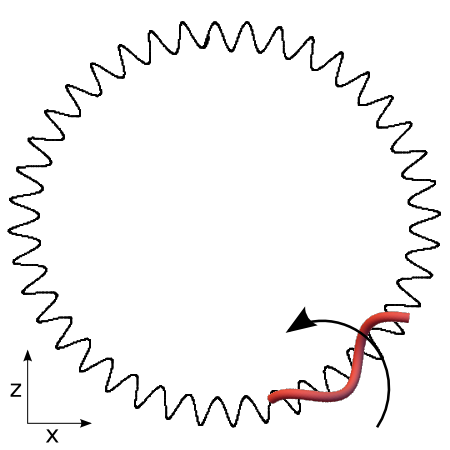}
\caption{drift: $\hat{y}$\label{diagramsteering1}}
\end{subfigure}\\
\begin{subfigure}{1\textwidth}
\includegraphics[width=1\textwidth]{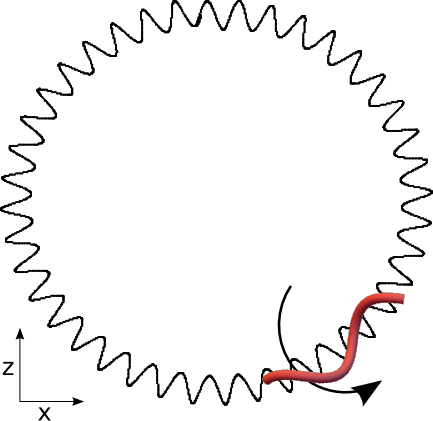}
\caption{drift: $-\hat{y}$\label{diagramsteering2}}
\end{subfigure}
\end{minipage}
\caption{Panels \textbf{(a-b)} show the trajectory of a helix driven at $700$ kHz with the perpendicular plane of the field rotated 
by $d \theta$ in the $\hat{y}$ direction at each time step. The helix drifts in the opposite direction of the displaced fluid the $\hat{y}$ direction. 
In general, the drift depends on the rotational velocity of the helix as it makes the turn. For angular rotations 
directed counterclockwise to the circular trajectory (panel \textbf{(c)}) the helix drifts in the $+\hat{y}$ direction, while 
for clockwise rotations the helix drifts in the $-\hat{y}$ direction (panel \textbf{(d)}).}
\label{fig:circular-trajectory}
\end{figure}

Finally, we demonstrate that smooth turns through curved or circular trajectories with or without thermal fluctuations are also possible as illustrated in Fig. \ref{fig:circular-trajectory}. 
The helix is driven at $700$ kHz with $MB=1.5\times10^{-18}$ N m
and the perpendicular plane of the rotating magnetic field is rotated by $d\theta=\pi/\delta t'$ for 
every time step where $\delta t'>f^{-1}$ (in the simulations $\delta t'=6.28f^{-1}$). 
Although the rotation of the perpendicular plane of the magnetic field is along the $\hat{y}$ direction, the resulting path shows a drift in 
the $y$ direction such that the intended circular path becomes helical. Left and right-handed helices were simulated with 
different combinations of clockwise and counterclockwise field rotations and directions of steering. In the deterministic simulations, 
we find that when the helix rotates such that it pushes the fluid inside the circular trajectory, 
the helix drifts in the opposite direction of the displaced fluid (cf. the diagram in 
Figs. \ref{diagramsteering1} and \ref{diagramsteering2})\footnote{A vector field plot of the fluid velocities can be found in the SI.}. 
In the presence of thermal fluctuations, this drift is totally overcome by Brownian motion since the P\'eclet number $\rm{Pe_T}=8$ in the $z$-direction.

\section*{Summary and conclusions}

In conclusion, we have shown that nanoscale helical propellers can be used for controlled motion and steering at high P\'eclet numbers.
The propulsion velocities and Pe of the propellers can be conveniently adjusted through the magnetic field frequencies for frequencies that are less than the step-out value,
which can be determined from the ratio of the magnetic interaction to drag as $f_{\rm SO}=MB/\kappa_\parallel$. 
Extending the range for higher propulsion velocities is possible by increasing both the field frequency and $MB$ or reducing drag. Compared with the fastest artificial controllable chiral micropropellers of $\sim 150$ nm in size, which have a maximum propulsion speed of $\sim40$ $\mu$m/s \cite{ghosh2009controlled,pak2011high}, our results show that nanoscale propellers of $\sim30$ nm in size can be propelled much faster with $\sim10^{-3}$ m/s with full control.
With these propulsion speeds, it is possible to attain ${\rm Pe_T}>100$, allowing spatial and temporal control 
of the motion {\it in the presence of thermal fluctuations}.

\subsection*{Computational model}

In the limit of $\rm{Re}=0$, the corresponding Stokes equations for particles of arbitrary shapes can be numerically solved and the friction matrix 
in Eq. \ref{eq1} unraveled.
In our recent work on helical particles
we have, however, shown that the Stokes approach is not sufficient for quantitative accuracy in the case of externally driven 
nanohelices \cite{Alcanzare2017}. To this end, we employ the fluctuating lattice-Boltzmann - Molecular Dynamics (LBMD) method of Ref. [\citenum{Ollila2011}].
It incorporates full Navier-Stokes hydrodynamics with consistent thermal fluctuations and a coupling of the fluid to extended MD particles of arbitrary shapes. 
The fluctuating LBMD has been extensively benchmarked for colloids and polymers \cite{ollila2013hydrodynamic,ollila2014biopolymer,mackay2014modeling}.
In the present case, 
the translational (rotational) Reynolds number for a helix of radius $R$ and length $L$ can be shown
\cite{Alcanzare2017} to be $\text{Re}_\text{T}=\rho v L/\eta$ ($\text{Re}_\text{R}=\rho \omega R^2/\eta$),
where $\rho$, $v$, $\omega$ $\eta$ are the fluid density, propulsion velocity, angular velocity and fluid viscosity, 
respectively. In the simulations, the typical translational (rotational) Reynolds numbers are about $10^{-4}-10^{-5}$ ($10^{-4}-10^{-2}$). Details about the method are provided in the Supporting Information (SI).

\section*{Acknowledgements (not compulsory)}

This work was supported in part by the Academy of Finland through its Centres of Excellence Programme (2012-2017) under Project No. 251748 and
Aalto Energy Efficiency Research Programme. We acknowledge the computational resources provided by Aalto Science-IT project and CSC-IT.
The graphical representations in the Fig. \ref{figure1} were rendered using VMD \cite{humphrey1996vmd}.

\section*{Author contributions statement}

M.M.A, M.K. and T.A.N. conceived the problem and analyzed the results.  All authors co-wrote and reviewed the manuscript.

\end{document}